# Charon: A brief history of tides

Alyssa Rose Rhoden*[1], Helle L. Skjetne[1], Wade G. Henning[2], Terry A. Hurford[3], Kevin J. Walsh[1], S. A. Stern[1], C. B. Olkin[1], J. R. Spencer[1], H. A. Weaver[4], L. A. Young[1], K. Ennico[1] and the *New Horizons* Team

1 – Southwest Research Institute, Boulder, CO
2 – Department of Astronomy, University of Maryland – College Park, College Park, MD
3 – NASA Goddard Space Flight Center, Code 698, Greenbelt, MD
4 – Johns Hopkins Applied Physics Lab, Laurel, MD
\* - Previously published under the surname Sarid

Highlights:

- Charon's tectonic features are inconsistent with eccentricity tidal stresses.
- If Charon had an internal ocean, it froze out after orbit circularization.
- Charon's fractures lack the equatorial symmetry of most tidal stress patterns.

Tides, interiors, and fractures on Charon

Corresponding author:	Alyssa Rhoden
	Southwest Research Institute
	1050 Walnut St. #300
	Boulder, CO 80302
	Alyssa@boulder.swri.edu

**Abstract:**

In 2015, the *New Horizons* spacecraft flew past Pluto and its moon Charon, providing the first clear look at Charon's surface. *New Horizons* images revealed an ancient surface, a large, intricate canyon system, and many fractures, among other geologic features. Here, we assess whether tidal stresses played a significant role in the formation of Charon's tensile fractures. Although presently in a circular orbit, most scenarios for Charon's orbital evolution include an eccentric orbit for some period of time and possibly an internal ocean. Past work has shown that these conditions could have generated stresses comparable in magnitude to other tidally-fractured moons, such as Europa and Enceladus. However, we find no correlation between observed fracture orientations and those predicted to form due to eccentricity-driven tidal stress. It, thus, seems more likely that Charon's orbit circularized before its ocean froze, and that either tidal stresses alone were insufficient to fracture the surface or subsequent resurfacing removed these ancient fractures.

**Plain language summary:**

Charon's surface displays many fractures that appear to have formed when the surface was pulled apart. The source of stress that created these fractures is currently debated. We test the hypothesis that, in the past, Charon had a slightly elliptical orbit, which would have stressed the surface, potentially creating the observed fractures. We find that this model provides a poor match to the observed fractures over a wide range of parameter values. These findings suggest that Charon's orbit was circular for most of its history, and that its ocean froze out after the orbit became circular. The observed fractures were likely formed as part of the freezing process, which has been suggested for other icy moons with canyons and fracture systems similar to those seen on Charon, although the details of this process are not well understood.

1. Introduction

In 2015, the *New Horizons* spacecraft flew past Pluto and its four moons, revealing the surfaces of these bodies for the first time. Charon's surface displays craters, tectonic features including large chasmata, raised features interpreted as mountains, and evidence of resurfacing from viscous, possibly cryovolcanic, flows (Beyer et al., 2017; Robbins et al., 2019). The cratering record on Charon suggests a ~4 Byr old surface (Beyer et al., 2017 and references therein; Singer et al., 2019), although it does not appear as heavily cratered as similarly aged bodies in the inner solar system. This disparity has been attributed to the lower flux of impactors and lower relative impact speeds within the Kuiper Belt. It has also been suggested that the southern portion of the encounter hemisphere is younger than the northern portion (using the informal feature names, Vulcan Planum is younger than Oz Terra), but both regions are billions of years old and the error bars allow for the possibility that they are the same age (Singer et al., 2019). Within the images obtained by *New Horizons*, it appears that craters have not been modified by tectonic features, which suggests that tectonic activity occurred early in Charon's history and was overprinted by craters (Singer et al., 2019).

It is generally accepted that a giant impact onto early Pluto resulted in Charon's formation (Canup, 2005; Desch et al., 2015). However, the details of the post-impact physical processes and orbital evolution of Charon are still largely unknown. The picture is even more complicated by the orbits of the other small moons of Pluto, which have not been successfully reproduced by any model of the system's evolution, to date (e.g., Walsh and Levison, 2015). Charon's orbit is currently circular and tidally-locked to Pluto, so it almost certainly underwent significant orbital evolution. Charon may have experienced an epoch of high eccentricity (e.g. Cheng et al., 2014) that could have driven significant internal heating and stress due to tides. Charon may also have possessed a subsurface ocean in its past (Desch et al., 2105; Desch and Neveu, 2017; Malamud et al., 2017, and references therein).

Rhoden et al. (2015) determined that an internal ocean would have made Charon much more responsive to the tidal deformation caused by eccentricity, particularly if Charon were evolving from a closer orbit to Pluto. Assuming a higher past eccentricity, across a broad range of interior structure parameters, stress magnitudes with an ocean approached those on Europa and Enceladus, which host tectonic features that have been linked to eccentricity-driven tidal stress (e.g. Rhoden et al., 2010; Patthoff and Kattenhorn, 2011). If fractures formed as a result of eccentricity-driven stresses, the orientations of fractures should vary across the surface in a specific pattern. Rhoden et al. (2015) made predictions as to those fracture patterns using pre-encounter assumptions about Charon's size, mass, and rock fraction.

An additional mechanism that has been proposed to explain Charon's tectonism is volume expansion due to freezing of an internal ocean (Moore et al., 2016; Malamud et al., 2017; Desch and Neveu, 2017). Stresses associated with freezing are isotropic, i.e. they have no preferred orientation. When combined with other sources of stress that do have preferred orientations, freezing stresses can amplify that pattern even if the magnitude of the e.g., tidal stress is low (Nimmo, 2004). Hence, if the ocean were freezing while Charon maintained an eccentric orbit, the fracture patterns should still correspond to the eccentricity pattern. Alternatively, Charon's orbit could have circularized and synchronized before the ocean froze. In that case, the extensional tectonic features that formed in response to the change in volume would not carry a signature of eccentricity.

Here, we update the eccentricity-driven fracture study of Rhoden et al. (2015) using post-encounter values for Charon's physical properties and compare our predicted fracture orientations with the observed fractures on Charon. We find that, over a wide range of conditions, eccentricity-driven tidal stresses do not produce a good match to the observed fractures, implying that Charon's tectonic activity was driven by a different process. Furthermore, the lack of eccentricity-driven tidal fractures within Charon's geologic record strongly suggests that, over the age of the current surface, tidal stress magnitudes were never high enough to induce failure. That also requires Charon's orbit to have circularized before the ocean froze (if it indeed had an ocean at all). Otherwise, the large stresses from freezing should have amplified the tidal stresses and created fractures. These results provide constraints on the orbital evolution of Pluto-Charon.

2. Methods

We began by selecting the geologic features that Robbins et al. (2019) classified as tensile fractures, such as grooves and graben (Figure 1). We recorded the latitude and longitude at the midpoint of each continuous feature. We then measured the orientation (or azimuth) of each feature, where 0° is north and the azimuth increases clockwise. Straight features were assigned an orientation confidence of 1, which comprise 78% of the data set. For features that did not display a consistent orientation (e.g. curved or jagged features), we measured the general trend of the feature's orientation and assigned the measurement a confidence of 2 or 3), which was determined subjectively (see Supplementary Online Material for feature data.

We then generated predictions of fracture orientations that would result from eccentricity-driven tidal stress from a presumed past epoch in which Charon's orbit was not circular. The past eccentricity of Charon is not well constrained, and its semimajor axis has changed with time. However, the stress patterns associated with eccentricity do not change with either of these quantities, so our results are not dependent upon the exact values we chose. In contrast, stress magnitudes scale linearly with eccentricity and with semimajor axis to the −1/3 (e.g., Kattenhorn and Hurford, 2009). We picked an eccentricity value of 0.01, which is the current eccentricity of Europa, and used Charon's current semimajor axis.

We then selected a suite of interior structures that are consistent with the mean radius and density of Charon from *New Horizons* measurements (Nimmo et al., 2017): 606.0 ± 1 km and 1701 ± 33 kg/m$^3$, respectively. We note that McKinnon et al. (2017) cites Nimmo et al. (2017) but reports a different density value. The reason for this discrepancy is unclear, but the density difference is too small to affect our results. All of the interior structures we tested include a rocky interior (core + mantle), a liquid water ocean, and an ice shell that is separated into a brittle portion atop a ductile portion. We assumed a constant hydrosphere (ice + liquid water) thickness of 230 km and tailored the rocky interior parameters to match the total mass of 1.5857x10$^{21}$ kg. The thickness of the hydrosphere is not well constrained; estimates depend on other uncertain parameters such as the density and porosity of the rocky interior, the proportion of rock to ice, and the degree to which Charon's shape is hydrostatic (McKinnon et al., 2017). The parameter values we have chosen result in an interior structure that is consistent with estimates from other models in that both the rocky interior and hydrosphere are hundreds of km thick and the rocky interior is thicker than the hydrosphere (McKinnon et al., 2017; Nimmo et al., 2017).

Parameter values held constant across all interior structures are provided in Table 1. Given the few constraints on Charon's interior, both currently and during its evolution, we only varied parameters that our past work (Rhoden et al., 2015, 2017) has shown to significantly contribute to the magnitude and orientation of tidal stress. Specifically, we varied the total thickness of the ice shell (and corresponding ocean thickness) from 30 km to 200 km and the viscosity of the ductile portion of the ice shell from $10^{13}$ Pa*s to $10^{15}$ Pa*s. Estimates of global extension due to freezing of an ocean suggest a 30 km ocean (Beyer et al., 2017), which gives us our lower bound. We explore oceans as thick as 200 km, corresponding to a 30 km ice shell, which may represent past conditions; thinner ice shells may be less likely to experience convection, which would make our 2-layer ice shell model less applicable (see Discussion). The ductile ice shell viscosities we explore span a range of values that are responsive to forcing at the orbital period, with the minimum value corresponding to the melting-point viscosity. Parameter values for each interior structure are given in Table 1.

To compute tidal stresses in a rheologically-layered body, we follow the approach of Jara-Orue and Vermeersen (2011), in which we calculate the responses of individual layers and layer interfaces to tidal forcing. We have previously implemented this methodology in studies of Mimas (Rhoden et al., 2017), pre-encounter Charon (Rhoden et al., 2015), and Enceladus (Rhoden et al., 2019). The equations were derived for a five-layer body in which the core and ocean layers are assumed to be fluid. As in our previous work (Rhoden et al., 2015; Rhoden et al., 2017), we find that reducing the thickness of the innermost layer (i.e., the liquid core) to a radius of 10 km, and assigning it the same density as the overlying silicate layer, is equivalent to using a four-layer model with a solid silicate innermost layer. Additional details as to the formulation, validation, and assumptions can be found in Rhoden et al. (2015) and Rhoden et al. (2017). For reproducibility, we have included output from the model that is specific to this work in the SOM.

For each interior structure we tested, we compute the principal tidal stresses for a set of locations and times in the orbit. We use a latitude and longitude grid: from 75°N to 75°S in increments of 15° (neglecting the equator) and 0°W to 360°W in increments of 30°. We divide the orbit into 1° increments and calculate stress at each of these points in the orbit. We can use these values to construct a plot of the largest tensile stress through the orbit for each location (in other words, the principal stress that is most tensile through time). We assume that fractures form perpendicular to the direction of the largest tensile stress at the point when failure occurs. We do not adopt a particular failure threshold because the failure threshold of Charon's ice shell is unknown and because the tidal stress magnitudes would scale upward with either a higher eccentricity, shorter orbital period, or the addition of cooling stresses. Instead, we report all fracture orientations associated with increasing tensile stress; we make the assumption that failure would occur while tensile stress is increasing toward the peak daily value in a given region and not as the stress is decreasing back toward zero (see cartoon in Fig. 2). This is a simplifying assumption. It is possible that fatigue stresses would allow a new fracture to initiate during the decreasing portion of the stress curve. Also, fractures that initiated in one location could propagate into a region in which stress is decreasing but still exceeds the stress intensity factor at the crack tip, enabling it to propagate into a region where we would not predict it to have initiated (as discussed further, below).

Our method for predicting fracture orientations differs somewhat from the predictions in Rhoden et al. (2015). There, we assumed that fractures in a given region were a combination of fractures formed there and fractures that propagated through the region but formed elsewhere. This assumption makes sense when looking at Europa's fractures, which can span 100s to 1000s of km (e.g. Kattenhorn and Hurford, 2009). We, thus, reported all orientations for which tidal stresses decomposed along a fracture of that orientation would be greater than a particular stress threshold. However, based on the mapping in Robbins et al. (2019), Charon's fractures tend to be shorter, and their endpoints are not obscured by subsequent resurfacing as they are on Europa. Therefore, we are making the simplifying assumption that, if tidal stresses are the cause of Charon's fractures, the initial failure orientations should dominate the observed patterns, which leads us to the assumption that fractures should form predominantly at times when the stress is increasing; we discuss the impact of this assumption more in the Discussion.

3. Results

For each interior structure model, we determined the highest magnitude of tensile tidal stress across all locations and times. We refer to this value as the peak stress magnitude (PSM) and use it to evaluate the propensity for failure to occur (see Section 4). As shown in Table 2, with our assumed eccentricity of 0.01, the PSMs range from 3 to 49 kPa.

In Figure 3, we display the orientations of Charon's observed fractures, which were identified as having morphologies consistent with tensile fractures by Robbins et al. (2019). Line symbols are not reflective of the actual lengths of the features, and the orientations of the lines neglect map projection effects. Note that the latitude/longitude dimensions of the figure have been trimmed to the limits of the mapped features, and the longitude scale does not reflect the Charon's curvature toward high latitudes. A histogram (inset) shows a strong preference for fractures with orientations between 0 and 90°; the distribution is the same whether we include all fractures or only the highest confidence fractures. We measure orientation as 0° north, increasing clockwise, which means the observed fractures have orientations in the first quadrant. The pie chart shows the most prominent orientations in dark green and the orientations with approximately a factor of two fewer features per bin in light green.

To compare predictions with observations, we binned the observed features to match the latitude-longitude grid we used for the predictions. In Figure 4, we show the predicted orientations of fractures that would form in response to eccentricity-driven tidal stress as grey lines, along with the observed fractures (green lines). The predicted orientations appear as wedges, but they are actually individual line symbols representing each orientation that is associated with increasing tensile principal stress. Because stress curves tend to be smooth curves, as depicted in the cartoon in Figure 2, the orientations will cluster because they change slowly as the stress increases. In certain cases, the tidal stress increases and decreases more than once during an orbit, which creates disconnected wedges, such as at 30°S, 0°E. The inset histogram shows predicted orientations from all regions (grey) and orientations that would form when the stress is at least half of the PSM (purple); we expect that fractures are more likely to form at higher stresses. In Figure 4a, we used a thin ocean (30 km), thick ice shell (200 km), and relatively cool ductile ice layer ($10^{15}$ Pa*s). In Figure 4b, we used a thick ocean (200 km), thin ice shell (30 km), and warm ductile ice layer ($10^{13}$ Pa*s). The predictions are very similar for these end member cases, and we see little variation across all of the interior structures we tested.

Based on both the maps and the histograms, these interior models do not provide good matches between the predictions and orientations across all features. We have also looked regionally, in case a difference in age or activity has affected fracture formation or preservation. Histograms for each region are shown in Figure 5 for the thick ocean case (Fig. 4a), in which green again represents the observed orientations. The grey data include predicted orientations that correspond to any magnitude of stress, as long as the stress is tensile and increasing. However, it is more likely that fractures form when stress is high; the purple data include only orientations that would form when the stress is within 50% of the PSM. In the southwestern region (A) of the *New Horizons* image coverage, the predictions are in the correct quadrant, although the observed population trends more north-south than we predict, and there is a broader range of observed orientations. Limiting our predictions to higher stresses results in a more concentrated set of predictions, with orientations between 75° and 90°. Eastward of this region, the fractures are anticorrelated with the predictions, as is the northern region (B) that surrounds Serenity Chasma. The higher stress predictions in this region have less of a mismatch because the number of predicted fractures decreases for larger azimuths. However, it is still the case that the model only matches fractures that run roughly east-west.

4. Discussion

The trends of tidal stress magnitude versus interior structure we find for Charon (Table 2) are the same as other bodies we have evaluated, including our pre-encounter study of Charon (Rhoden et al., 2015). With a ductile ice viscosity of $10^{13}$ Pa*s, the PSM increases with increasing shell thickness because material with a viscosity of $10^{13}$ Pa*s is highly responsive to forcing at the orbital period. The thicker the layer, the more material there is to respond, leading to higher stress. When the ductile ice viscosity is $10^{14}$ Pa*s or $10^{15}$ Pa*s, the opposite happens; the PSM decreases as the shell thickens because a thinner shell can flex more easily in response to tidal forcing than a thicker shell. A warm, thick, low viscosity ductile layer can thus produce higher tidal stress magnitudes. However, it may not be a realistic representation of an ice shell.

The melting point viscosity of ice is likely near or above $10^{13}$ Pa*s (e.g. Collins et al., 2009). For an ice shell to maintain a constant, low viscosity through 10s of km of ice requires that the ice shell is warm and nearly isothermal, implying convection. Based on the parameters we have assumed for Charon, we find that convection is plausible (see Appendix A). However, tidal stress orientations, and specifically the symmetry in the predicted fracture patterns from tidal stress, would not be significantly different for a conductive model, so we can still draw conclusions about the role of tidal stress in the development of Charon's fractures even if our model has oversimplified the thermal profile of Charon. More sophisticated tidal stress models that can incorporate enough layers to mimic a conductive profile would be useful in evaluating more realistic ice shell rheologies in the absence of convection, but the 2-layer model should be sufficient for our purposes.

Models of the long-term, thermal-orbital evolution of Europa show that there are two types of stable end states for ice shells (Walker, 2017). We have modeled the case of a relatively thin brittle layer, but it is also possible for an ice shell to be almost entirely cold and brittle, with a thin convecting region at the base of the shell. Based on past work in which we varied the brittle layer thickness (e.g. Rhoden et al., 2015), we expect that thickening the brittle layer would reduce the tidal stress magnitudes for interior structures with low viscosity ductile ice layers. It has little effect on ice shells with ductile ice layers that have viscosities of $10^{14}$ Pa*s or above (Rhoden et al., 2015; Patthoff et al., 2019).

Using Charon's current orbital distance and an assumed eccentricity of 0.01, the tidal stress magnitudes we find for an ocean-bearing Charon are lower than the range of failure stresses implied by Europa's tidally-driven fractures (50 – 100 kPa) and for terrestrial sea ice (Dempsey et al., 1999) and much lower than the tensile failure strength of pure, water ice in laboratory, which exceeds 1 MPa (Schulson, 2006; Collins et al., 2009). Because tidal stresses scale with eccentricity, and to a lesser extent, the distance from the primary, a larger past eccentricity and/or closer orbit could have generated larger values that could induce failure. In contrast, if Charon's orbital eccentricity never reached 0.01, the eccentricity stresses would be lower than we find here. The addition of stress through freezing of the subsurface ocean would also generate large stresses that could amplify a tidal stress signature (Nimmo, 2004; Hemingway et al., 2019). Therefore, the small magnitudes of tidal stresses cannot, on their own, rule out the possibility of such fractures forming on Charon.

The fracture predictions from Rhoden et al. (2015) differ from our current predictions because, here, we did not include orientations of features that could have propagated through the region but formed elsewhere. Rather, we limited our orientation predictions to the initiation of new fractures, which we assume formed when the tidal stress was increasing. The 2015 predictions are mirrored in longitude across the sub-Pluto point as well as in latitude across the equator. The predicted orientations are mainly east-west, and the range of predicted orientations broaden closer to the equator and to the sub-Pluto point (see Rhoden et al., 2015). Assuming a lower failure threshold also allows a broader range of orientations. Although there are a few locations where the predicted orientations correlate well with the observed fractures, the overall pattern of observed fractures does not match the 2015 predictions or the ones presented here.

Hence, the simplifying assumptions in this work are not the cause of the misfit between the predictions and the observations.

Given the poor match between Charon's observed fractures and the predicted fracture orientations due to eccentricity, we conclude that the mapped fractures were not caused by eccentricity-driven tidal stress. We anticipate a similar challenge invoking tidal stresses from obliquity or physical libration, neither of which have been inferred to be non-zero from any measurements to date, due to the symmetries in their resulting stress fields, so we have left such investigations to future work. We note that there are no longitudes at which the predicted fracture orientations from eccentricity match the observed fractures in the northern hemisphere, which trend northeast-southwest. Non-synchronous rotation of an ice shell has been proposed for multiple ocean-bearing moons as a way of reconciling stress predictions with observed fracture patterns (e.g. Rhoden et al., 2010 for Europa; Patthoff and Kattenhorn, 2011 for Enceladus); longitudinal migration of features due to NSR would not improve the fit in this case. We have not evaluated a stress field that combines NSR stress with eccentricity-driven stress, which could affect the overall pattern of fracture orientations. However, the lack of north-south trending fractures on Charon is inconsistent with patterns expected from NSR for equatorial or midlatitudes (e.g. Kattenhorn and Hurford, 2009).

A translation in latitude could place more fractures in the southern hemisphere, where the predicted orientations are generally in the first quadrant, more similar to the observations. However, that would require a substantial polar wander event (e.g. ~60° of rotation), for which there is no independent evidence nor expectation. It would also imply that nearly all the fractures observed by New Horizons predate the rotation and nearly none formed as a result of, or after, the rotation. Fairly rapid polar wander of an ice shell over an ocean can itself generate stresses that are generally much larger than those from diurnal tides. The polar wander stress field can be complex (e.g. Matsuyama and Nimmo, 2008); it depends upon the locations of the initial and final poles, whether the rotation axis is aligned with the tidal axis, and the extent of despinning (or spin up) that occurs as the pole changes. The stress field is still largely symmetric, but the line of symmetry is a great circle related to the pole locations, so it traces out a curve in both latitude and longitude. Polar wander stresses form provinces of thrust and normal faults, while strike-slip faults form in a wide band between the poles. Although we have only observed a portion of Charon's surface, the lack of widespread strike-slip faulting and the relatively consistent orientation of fractures along a fairly wide range of longitudes and latitudes (i.e. a lack of identifiable symmetry along any great circle), makes polar wander an unlikely source of stress to explain Charon's observed fractures.

Pluto and Charon are now tidally locked, meaning they rotate synchronously with each other. Evolving to this state would have involved Charon receding from Pluto and despinning (i.e. its spin period slowing down) during its early evolution. Despinning and orbital recession generate tidal stresses because a body must adopt a new primary tidal shape. Stress patterns from these two effects, with a generic set of inputs for a Europa-like moon, are shown in Kattenhorn and Hurford (2009) and Matsuyama and Nimmo (2008), which also accounts for tidal deformation. An important difference between these tidal stresses and those caused by eccentricity is that they can continually increase until failure occurs. An example of this process was shown for Mars' moon, Phobos (Hurford et al, 2016), although in that case the orbital distance is decreasing so the stresses are inverted (i.e. tension instead of compression) from the pattern we would predict for Charon. With orbital recession and despinning stresses, multiple generations of fractures could form, relieving tidal stress, with interspersed periods of quiescence while stress reaccumulate.

The stress field associated with this evolution is the superposition of the collapse of the tidal bulges, due to a larger distance between Pluto and Charon, and the slowing of the spin rate. Orbital recession stresses mimic the eccentricity-driven stress field when a body is at apocenter. Because we have included all orientations associated with increasing tidal stress, the orientations allowed by recession are already represented in our predictions. When despinning is combined with recession stresses, the predicted fracture orientations would differ from an eccentricity stress field, but the symmetry across the equator would persist (e.g. Fig. 4 of Matsuyama and Nimmo, 2008). The observed fractures on Charon do not display this symmetry; fractures are oriented mainly northeast-southwest both north and south of the equator.

We conclude that either eccentricity-driven fractures never formed on Charon or that they formed extremely early in Charon's history and were obliterated by subsequent resurfacing processes. In either case, if there was an ocean within Charon, the orbit must have circularized before the ocean froze out. Otherwise, the combination of large cooling stresses and (small) tidal stresses should have enabled widespread fracturing with a pattern consistent with the eccentricity stress field. In addition, circularization must have occurred early enough that fractures caused by orbital recession and despinning stresses were not retained in the geologic record.

It seems plausible that circularization of Charon's orbit would have reduced tidal heating, eventually resulting in freezing of the ocean and a volume change that formed Serenity Chasma, although some additional process is required to explain its orientation and location. This hypothetical evolution is supported by Charon's global topography (Schenk et al., 2018), which implies limited heat flows, as compared with other bodies, and a limited lifetime for any potential ocean. Our results are consistent with either of the following scenarios: the circularization process happened early enough that tidally-driven fractures were overprinted by subsequent geologic activity or that the tidal stresses were never large enough to create fractures without cooling stresses to enhance them.

Saturn's moon, Tethys, may have undergone a similar history. These two moons are similar in size (mean radius of 531 km for Tethys and 603 km for Charon), although Charon's mean density is about twice that of Tethys. Both moons currently maintain circular orbits, lack evidence of eccentricity-driven fractures, and have large canyon systems. Viscous relaxation of craters on Tethys suggest a past epoch of high heat flows (White et al., 2017), and simulations of the thermal-orbital evolution of Saturn's moons imply periods of high eccentricity and the presence of an internal ocean in Tethys (Neveu and Rhoden, 2019). Both Charon and Tethys may have lost their oceans after circularization, leading to changes in volume that formed the observed chasms (Desch et al., 2015; Desch and Neveu, 2017; Malamud et al., 2017). The geologic records of both Charon (Beyer et al., 2017; Robbins et al., 2019) and Tethys (Ferguson et al., 2019) include fractures at a variety of orientations and sizes; not all fractures are obviously related to the formation of the main canyon system. Hence, we are left with two questions: 1) What governed the orientation of the main canyon systems on these two moons, given that the stresses associated with freezing an ocean are isotropic and 2) what process(es) led to the formation of other tectonic features on these moons?

While Tethys and Charon are the most well-studied examples, these unexplained, global-scale tectonic processes may be common on mid-sized planetary satellites. Of particular note, the Uranian mid-sized satellites (Oberon, Titania, Umbriel, Ariel) all possess similar, global-scale extensional grooves and graben (Croft & Soderblom, 1991). Analysis of the Uranian satellites is hampered by limited imaging data from *Voyager 2*. Nonetheless, this begs the question of whether there are common processes at play in this size-class of planetary satellites. If ocean freezing is the underlying driver of chasm formation and other tectonic activity on these moons, more modeling is needed to understand the mechanism that concentrates the orientations of the features. Some possibilities include preexisting (tidal?) fractures altering the otherwise isotropic stress field caused by cooling or ice shell thickness variations enhancing stress in certain regions, leading to failure that then further alters the stress field.

5. Conclusions

Even though Charon is expected to have had an eccentric orbit and undergone tidal recession and despinning, these events have not been preserved in Charon's tectonic record. The lack of tidal fractures implies that either tidal stresses were never high enough to produce fractures or Charon's geologic record was reset after the epoch of tidally-driven fracture formation. The most likely potential explanations are either Charon never had an ocean, so tidal stress magnitudes were even lower than we estimate here and freezing stresses were unavailable, or that Charon's orbit circularized before the ocean froze out. In that case, there would be no eccentricity or despinning stresses to combine with the freezing stresses and generate a distinct pattern. Given that the formation of Charon's chasm system has been attributed to the volume expansion caused by a freezing ocean, we favor the interpretation that Charon's orbit circularized early in its evolution, the lack of tidal heating contributed to ocean freezing, and the freezing ocean generated most of the tectonic features we observe, removing evidence of past tidally-driven fracturing. The mechanism by which Charon's observed fractures formed at their particular orientations, thus, remains an open question. Fully coupling the thermal, tidal, and orbital evolution of Pluto and Charon, while accounting for the constraints implied by the lack of tidally-driven fractures on Charon and the orbits of the smaller moons, may be critical to understanding the formation of this complex system.

Acknowledgements: This work was supported by the *New Horizons* mission and benefitted from discussions with members of the Ocean Worlds Cooperative. The results reported in this paper can be reproduced using the parameter values given in the text, tables, and supporting online materials, along with the equations and descriptions provided in the cited literature. The authors have no competing interests to report.

Appendix A:

As discussed in Section 4, the lack of any discernable spatial pattern in Charon's observed fractures makes a tidal stress origin unlikely regardless of the specific parameter values or model assumptions we employ. However, for completeness, we briefly test the plausibility that a sublayer of Charon's ice shell will be in a convective regime, as opposed to possessing a purely conductive thermal profile with depth. Our two-layer ice shell model is a more accurate representation of a convective ice shell than a conductive one.

We compute the Rayleigh number for a geophysical convective layer as a function of thermal flux via (O'Connell and Hager, 1980):

$$Ra = (\alpha\, g\, \rho\, d^4\, q_{BL}\,) \,/\, (\,\eta(T)\, \kappa\, k_{therm}\,), \tag{A1}$$

where $\alpha$ is the coefficient of thermal expansivity, $g$ the local gravitational acceleration, $\rho$ the local density, $q_{BL}$ the thermal heat flux through the top of the layer under consideration, $k_{therm}$ the thermal conductivity, and $\eta(T)$ the local viscosity as a function of temperature. The thermal diffusivity $\kappa$ is computed as $\kappa = k_{therm} / \rho\, C_p$, where $C_p$ is the specific heat at constant pressure. All of these material properties for ice are subject to uncertainty and often exhibit significant variation with temperature, as well as ice porosity and the presence of impurities.

For a basic analysis, we select $C_p = $ ~2050 J Kg$^{-1}$ K$^{-1}$ (Dorsey, 1940), $k_{therm} = $ ~2.3 W m$^{-1}$ K$^{-1}$ (e.g., Hobbs, 2010; Rathbun et al., 1998), and $\alpha = $ ~5×10$^{-5}$ (e.g., Eisenberg and Kauzmann, 2005; Röttger et al., 1994), each roughly appropriate for a convective shell at approximately 260−270 K. For this study, we assumed viscosities in a typical fixed range of 10$^{13}$ to 10$^{15}$ Pa*s, as well as ice densities near 1000 kg m$^{-3}$. The gravitational acceleration at the surface of Charon is 0.29 m s$^{-2}$. The thickness of the layer under consideration, $d$, is varied from ~ 200 km to 30 km (25 km ductile); thinner shells have greater difficulty convecting.

For $q_{BL}$, we first neglect any heat generated by possible tidal heating, and just consider the possibility of convection in Charon's shell due to radiogenic heating in Charon's silicate/iron core alone. As a low bound, Earth's upper-mantle volumetric radiogenic heat rate, at present, is approximately 10$^{-8}$ W m$^{-3}$ (e.g., Korenaga, 2008). Radiogenic rates are generally higher in the upper mantle and crust, as well as within the bulk silicate/iron component of chondritic material at $1.8 \times 10^{-8}$ W m$^{-3}$ (Multhaup and Spohn, 2007; Schubert et al., 2007). Radiogenic heating rates will also have been higher in the past (~10× higher ~4.5 Gyr ago), such as when Charon's fractures may have formed. The nominal values above therefore represent somewhat conservative under-estimates. Using a silicate-iron total core radius of ~377 km for Charon, calculating volume and blending with the chondritic rate above, leads to $4.0 \times 10^9$ W of radiogenic heat at present. Dividing by the surface area of Charon leads to $q_{BL} \sim 8.7 \times 10^{-4}$ W m$^{-2}$.

Using these values, we compute Rayleigh numbers via equation A1. At $\eta = 1 \times 10^{14}$ Pa*s, solutions range from approximately $1.8 \times 10^4$ ($d = 25$ km), to $7.6 \times 10^7$ ($d = 200$ km), and vary linearly with any assumed variation in viscosity. Such values are far in excess of estimates for the critical Rayleigh number required for stable maintenance of ongoing convection (e.g., $Ra > 1100-1700$, Chandrasekhar, 1961; O'Connell and Hager, 1980), as well as partly overlapping the higher thresholds required for the onset of convection in an unstable system with hysteresis and detailed rheology. See e.g., the more comprehensive methods in McKinnon (1999) and Barr & McKinnon (2007), which focus on basal Raleigh number $Ra_b$, for convective *onset* in a non-Newtonian system with strong viscosity contrasts, with higher thresholds ($Ra_{b\,cr} \sim 1 \times 10^5 - 1 \times 10^7$, see e.g., Table 2 of Barr and Showman, 2009), instead of bulk adiabatic core $Ra$ computations for isoviscous maintenance, as above.

For nominal ice material property values above, maintenance of ongoing convection for Charon becomes significantly questionable below sublayer thicknesses of approximately 15 km, if given also the conservative estimate of radiogenic-only heating from below. If heating is increased by a factor of ten (e.g., due to tidal heating in the past), a thickness of 9 km might still maintain convection. If heating is increased by 100× (tides plus higher radionuclide heating in the past, or any equivalent variation of governing parameters), convection might plausibly be maintained down to a ~5 km sublayer thickness, well below the range of shell dimensions assumed in this study. At such heat rates, even at our high-end viscosity of $\eta = 1 \times 10^{15}$ Pa*s, $Ra$ values for our low-end sublayer thickness $d = 25$ km, almost fully overlap the typical range of $Ra_{b\,cr}$ values for high-viscosity contrast convective initiation. Past tidal heating may have even exceeded such levels, depending in particular on Charon's past eccentricity (e.g., Saxena et al., 2009). The exact details of shell convection are complex and require finite element analysis (e.g., Showman and Han, 2005; Barr, 2008). However, this computation serves to support a general assumption of a convective-like thermal structure for Charon as used in this work.

Table 1. Parameters held constant across all models

| | | |
|---|---|---|
| Orbital period | | 6.38723 d |
| Surface gravity | | 0.279 m/s$^2$ |
| Radius | | 606.0 km |
| Mass | | 1.5857x10$^{21}$ kg |
| Eccentricity | | 0.01 (assumed) |
| Core | Thickness | 10 km |
| | Rigidity | 0 Pa (liquid) |
| | Density | 3938.2 kg/m$^3$ |
| | Viscosity | 0 Pa*s (liquid) |
| Mantle | Thickness | 366 km |
| | Rigidity | 5x10$^{10}$ Pa |
| | Density | 3938.2 kg/m$^3$ |
| | Viscosity | 10$^{19}$ Pa*s |
| Ocean density | | 1000 kg/m$^3$ |
| Ice | Rigidity | 3.487x10$^9$ Pa |
| | Density | 999 kg/m$^3$ |
| | Brittle layer thickness | 5 km |
| | Brittle layer viscosity | 10$^{21}$ Pa*s |

Table 2. Interior models and resulting stress magnitudes

| Ice shell thickness (km) | Ductile layer viscosity (Pa*s) | Peak stress magnitude (kPa) |
|---|---|---|
| 30  | 1.00E+13 | 32 |
| 60  | 1.00E+13 | 37 |
| 100 | 1.00E+13 | 44 |
| 200 | 1.00E+13 | 49 |
| 30  | 1.00E+14 | 27 |
| 60  | 1.00E+14 | 24 |
| 100 | 1.00E+14 | 19 |
| 200 | 1.00E+14 | 11 |
| 30  | 1.00E+15 | 13 |
| 60  | 1.00E+15 | 9 |
| 100 | 1.00E+15 | 7 |
| 200 | 1.00E+15 | 4 |

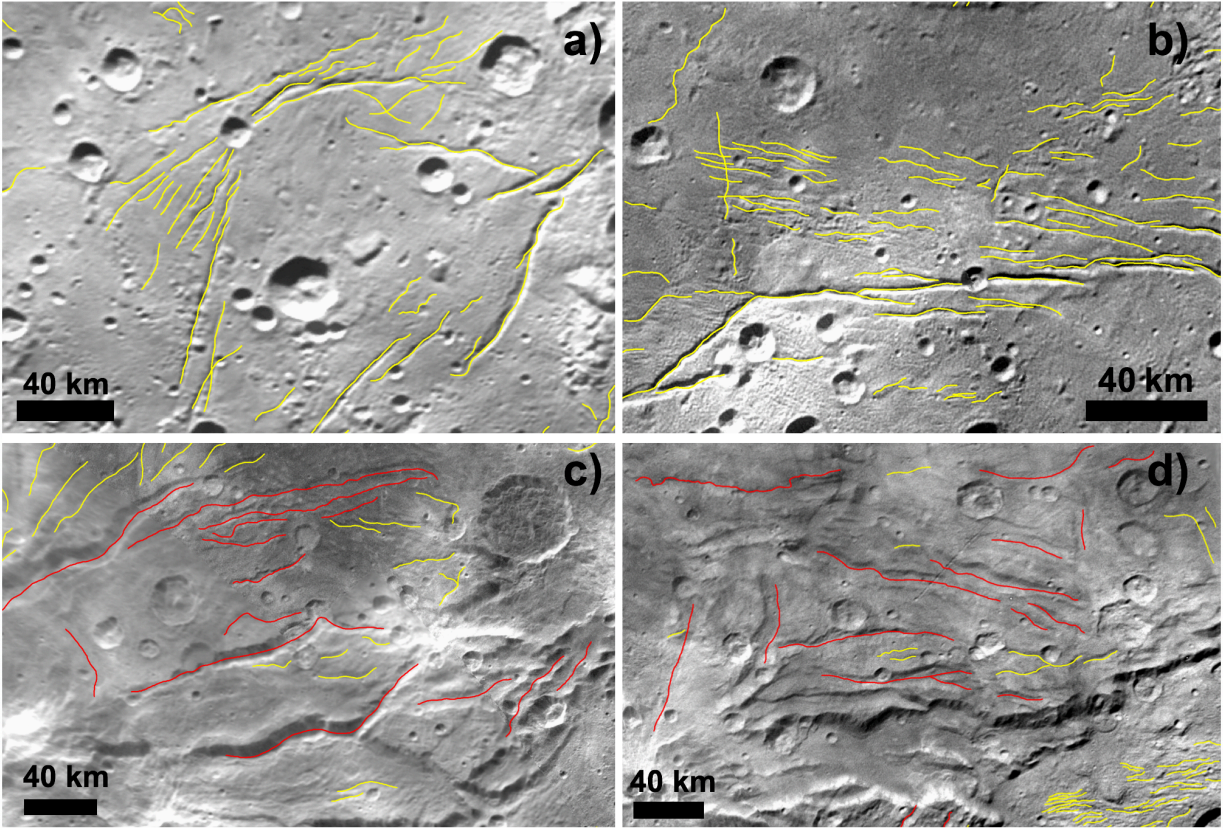

**Figure 1:** Features on Charon, categorized as tensile by Robbins et al. (2019), include grooves (yellow; panels a and b) and graben (red, panels c and d). We measured the orientations of these features for comparison with predictions from tidal models. Center points of each panel are located at a) 153°E, 15°S, b) -174°W, 5°N, c) 153°E, 16°N, and d) -170°W, 27°N.

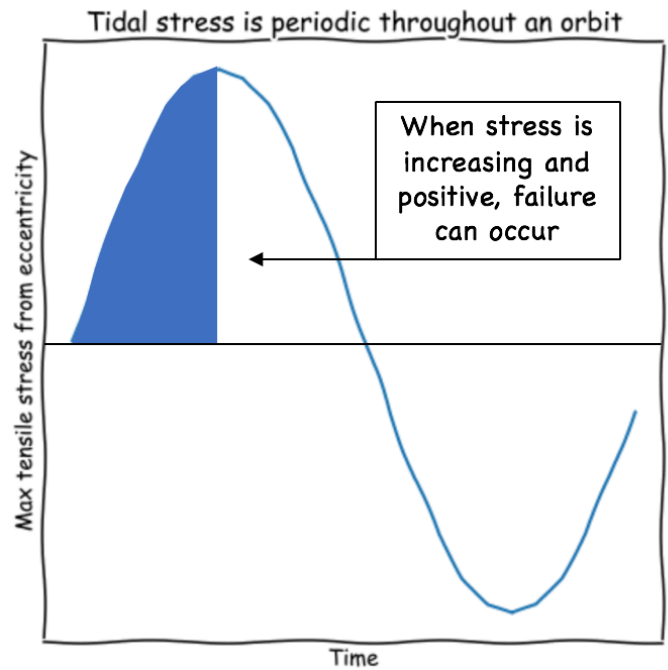

**Figure 2:** We calculate the most tensile of the two principal tidal stresses (y-axis) and its orientation (x-axis) as Charon moves through its orbit. We assume that fracture formation can occur only when the stress is positive (tensile) and increasing, as indicated by the shaded region in this cartoon. We also assume that fractures form perpendicular to the direction of the max tensile stress, which is typical for Mode I (opening) cracks.

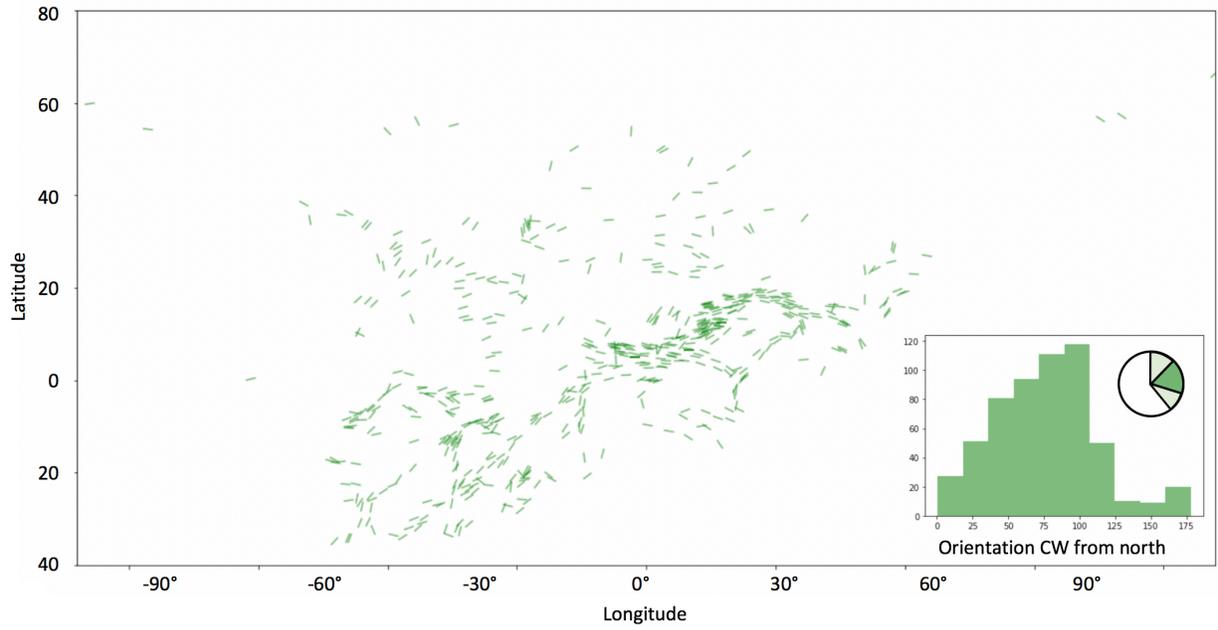

**Figure 3:** Green lines represent the location of the midpoint of each feature and its measured orientation. Features were initially mapped and categorized by Robbins et al. (2019). A histogram of the observed orientations (inset) shows that most features are oriented in the first quadrant (N to NE to E), as represented graphically with the pie chart.

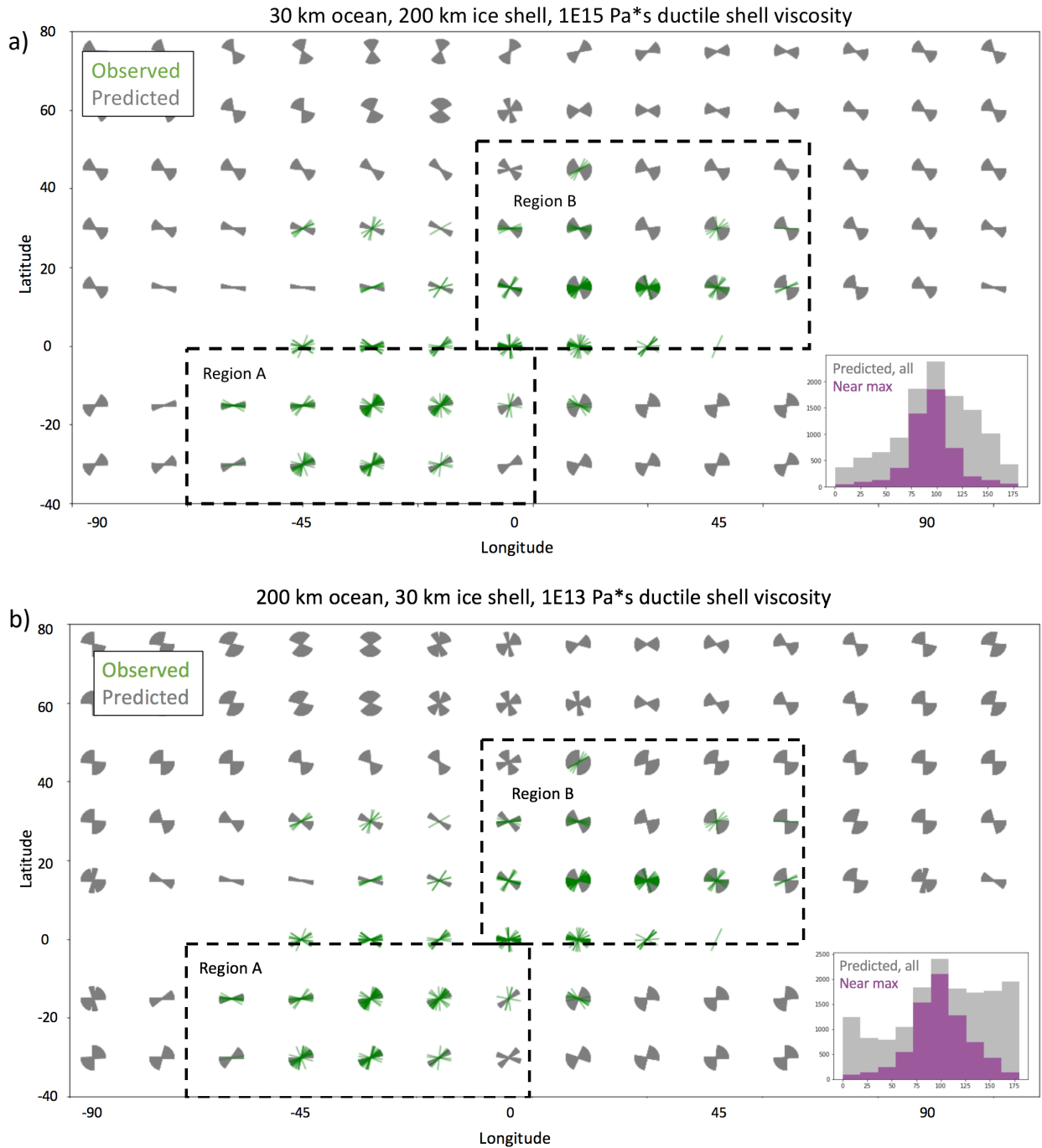

**Figure 4:** We binned the observations to match the latitude-longitude grid of our predictions and overlaid the observed orientations (green) onto the predicted orientations (individual grey lines, which tend to cluster into wedges). The inset histogram shows that, across all regions, eccentricity tidal stresses would tend to produce east to southeast trending faults. The predictions are more strongly peaked at east-west orientations if we include only those that form when the stress is at least half the peak stress magnitude. These predictions only partially overlap the observations (histogram shown in Fig. 3). We find similar results for a (a) thick, cool ice shell and (b) thin, warm ice shell.

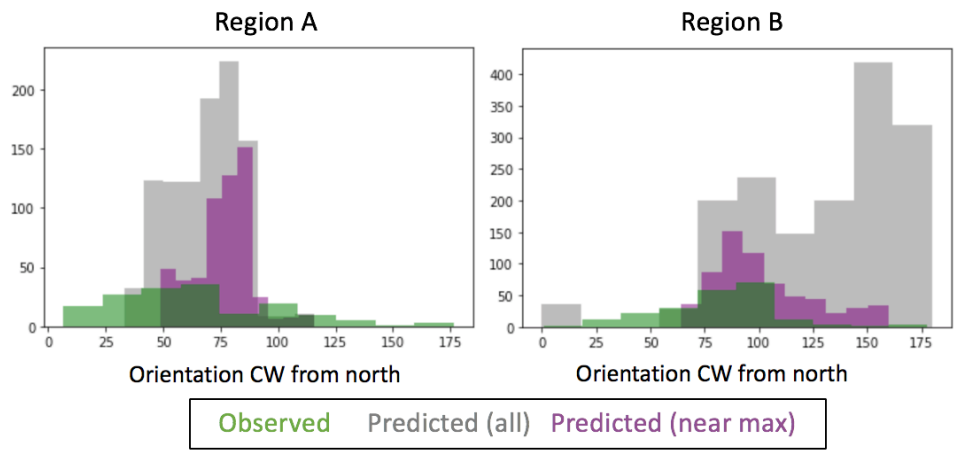

200 km ocean, 30 km ice shell, 1E13 Pa*s ductile shell viscosity

**Figure 5:** Regional histograms show that no areas are well matched by the tidal stress predictions. Here, we have shown the predictions assuming fractures can form at any magnitude (grey) or when the stress is within 50% of the peak stress magnitude (purple).